\documentclass{PoS}

\usepackage{amssymb,amsmath}
\usepackage{braket}
\usepackage{caption}
\usepackage{subcaption}
\usepackage{feynmf}
\usepackage{multirow}
\usepackage{slashed}
\usepackage{float}
\usepackage{adjustbox}

\title{BSM Kaon Mixing at the Physical Point}

\ShortTitle{BSM Kaon Mixing at the Physical Point}

\author{\speaker{Julia Kettle}\\
		School of Physics and Astronomy,University of Edinburgh, EH9 3FD, Edinburgh, UK \\
        E-mail: \email{J.R.Kettle-2@sms.ed.ac.uk}}

\author{Peter Boyle\\
        School of Physics and Astronomy,University of Edinburgh, EH9 3FD, Edinburgh, UK \\
        E-mail: \email{paboyle@ph.ed.ac.uk}}

 \author{Nicolas Garron\\
 	Dept. of Mathematical Sciences, University of Liverpool, L69 3BX, Liverpool, UK\\
 	E-mail: \email{Nicolas.Garron@liverpool.ac.uk}}
  \author{Tobias Tsang\\
 	School of Physics and Astronomy, University of Edinburgh, EH9 3FD,
 	Edinburgh, UK \\
 	E-mail: \email{}}
 \author{The RBC and UKQCD collaborations}

\abstract{We present new preliminary results for bag parameters and ratios of the BSM kaon mixing operators measured at the physical point. The results are obtained from simulations of domain wall fermion QCD with 2+1 flavours with an Iwasaki gauge, and now include pion and kaon masses very close to the physical point. We compare these results to our collaboration's previous results obtained from heavier simulated quarks.}

\FullConference{34th annual International Symposium on Lattice Field Theory\\
		24-30 July 2016\\
		University of Southampton, UK}

\begin{document}

\section{Introduction}

The motivation to study neutral kaon mixing comes from the role it plays in indirect CP violation observed in kaon decays. The degree of indirect CP violation is a well known, experimentally measured quantity \cite{Agashe:2014kda}, which can be expressed as a simple product of the kaon mixing bag parameter, Wilson coefficients and a function of CKM matrix elements and other known quantities. Thus precise lattice calculations of kaon mixing help to constrain the CKM matrix and beyond the standard model (BSM) theories (where new CP violating phases can can appear.) In the standard model (SM), kaon mixing is mediated by the W boson and the leading order contributions are given by the box diagrams. An operator product expansion (OPE) separates the short and long distance contributions. The  short distance part is given by the matrix element $\braket{\bar{K}^0|O_1|K^0}$, where $O_1$ is a $\Delta S = 2$ effective four-quark operator, given in equation \ref{eq:ops}, and can be found non-perturbatively using lattice simulations.

\begin{figure}[H]
	\centering
	\begin{subfigure}{0.45\linewidth}
		\centering
	\begin{fmffile}{KaonMixing} 
		\fmfframe(10,10)(10,10){ 
			\begin{fmfgraph*}(90,35) 
				\fmfleft{i1,i2} 
				\fmfright{o1,o2} 
				\fmflabel{$d$}{i2} 
				\fmflabel{$d$}{o1} 
				\fmflabel{$s$}{i1} 
				\fmflabel{$s$}{o2} 
				\fmf{fermion}{i1,v1} 
				\fmf{fermion,tension=.5,label=$u,,c,,t$,1.side=right}{v1,v3} 
				\fmf{fermion}{v3,o1} 
				\fmf{fermion}{o2,v4} 
				\fmf{fermion,tension=.5,label=$u,,c,,t$,1.side=right}{v4,v2} 
				\fmf{fermion}{v2,i2} 
				\fmf{photon,tension=0,label=$W$,1.side=left}{v1,v2} 
				\fmf{photon,tension=0,label=$W$,1.side=left}{v3,v4} 
				\fmfdotn{v}{4} 
			\end{fmfgraph*} 
		}        
	\end{fmffile}
	\caption{Example of a Box Diagram} 
	\end{subfigure}
    \begin{subfigure}{0.45\linewidth}
    	\centering
	\begin{fmffile}{4qOp}
		\fmfframe(8,8)(8,8){
			\begin{fmfgraph*}(70,45)
				\fmfleft{i1,i2}
				\fmfright{o1,o2}
				\fmflabel{$d$}{i2}
				\fmflabel{$d$}{o1}
				\fmflabel{$s$}{i1}
				\fmflabel{$s$}{o2}
				\fmf{fermion}{i1,v}
				\fmf{fermion}{i2,v}
				\fmf{fermion}{v,o1}
				\fmf{fermion}{v,o2}
			\end{fmfgraph*}
		}
	\end{fmffile}
\caption{Effective four-quark operator}
\end{subfigure} 
\end{figure} 
\vspace{-0.5cm}
Beyond the standard model, where the mediating particle is not restricted to the W boson, a range of operators become available as more dirac and colour structures are possible. A basis of 5 $\Delta S =2$ effective BSM operators can be defined \cite{Gabbiani:1996hi},
\vspace{-1cm}

\begin{equation}
\centering
\begin{split}
O_1 &=  [\bar{s}_a \gamma_{\mu} (1-\gamma_5) d_a][\bar{s}_b \gamma_{\mu} (1-\gamma_5) d_b]  \\
O_2 &=  [\bar{s}_a (1-\gamma_5) d_a][\bar{s}_b (1-\gamma_5) d_b] \qquad \qquad
O_3 =  [\bar{s}_a (1-\gamma_5) d_b][\bar{s}_b (1-\gamma_5) d_a]\\
O_4 &=  [\bar{s}_a (1-\gamma_5) d_a][\bar{s}_b (1+\gamma_5) d_b]\qquad \qquad
O_5 =  [\bar{s}_a (1-\gamma_5) d_b][\bar{s}_b (1+\gamma_5) d_a] \\
\end{split}
\label{eq:ops}
\end{equation}
\vspace{-0.75cm}

where a and b are colour indices, and dirac indices are not shown. This basis is referred to as the SUSY basis, but is model independent.
By calculating the matrix elements of these operators we can place boundaries on BSM theories. 

\section{Simulations} 

The simulations are of domain wall fermion QCD with Iwasaki gauge fields.  We present new results from two ensemble sets; the $48^3$ and $64^3$ physical point, details of which can be found in ref. \cite{1411.7017}. Alongside the physical point ensembles, in our analysis we include results from the ensemble sets $24^3$ and $32^3$ in our analysis. Details of our $32^3$ can be found in ref. \cite{Aoki:2010dy}, however $24^3$ has been recalculated with a physical strange mass. A summary of simulation parameters for the ensembles included is shown in table \ref{tab:simparams}. We have also performed calculations on a finer lattice, referred to as $48^3$ fine, but the NPR calculation is still ongoing and so we have not included those results here, as they cannot yet be included in our analysis. This will be included in a later journal publication writing up this work, and hence our present results should be viewed as preliminary.

\begin{table}[H]
	\centering
	\begin{tabular}{c | c  c | c c}
	& 48 & 64 & 24  & 32 \\
	\hline
	$(L/a)^3\times (T/a)$ & $48^3 \times 96$ & $64^3 \times 128$ & $24^3 \times 64$ &  $32^3 \times 64$\\
	$a^{-1} (\mathrm{GeV})$ & 1.7295(38) & 2.3586(70) & 1.7848(50) & 2.3833(86) \\
	$am_l$ & 0.00078 & 0.000678 & 0.005,\hspace{0.01cm} 0.01 & 0.004,\hspace{0.01cm} 0.006,\hspace{0.01cm} 0.008 \\
	$am_{s,val}$ & 0.0362 & 0.02661 & 0.03224 &  0.0250, \hspace{0.01cm} 0.0300 \\
	$am_{s,val}^{phys}$ & 0.03580 & 0.02359 & 0.03224  & 0.02477\\ 
	\end{tabular}
\caption{A summary of simulation parameters for each ensemble.}
\label{tab:simparams}
\end{table}


\section{Measurements}

The Bag Parameter is defined as the matrix elements normalised by their VSA values, expressed as,

\begin{minipage}{0.5\linewidth}
	\begin{equation}
	B_1(\mu) = \frac{\braket{\bar{K}|O_1(\mu)|K}}{\frac{8}{3}m_K^2 f_K^2}
	\end{equation}
\end{minipage}
\begin{minipage}{0.5\linewidth}
	\begin{equation}
	B_i(\mu) = \frac{\braket{\bar{K}|O_i(\mu)|K}}{N_i m_K^2 f_K^2 (\frac{m_K}{m_u(\mu)+m_s(\mu)})^2}
	\end{equation}
\end{minipage} for the SM and BSM case respectively, where $N_i=(−\frac{5}{3}, \frac{1}{3},2, \frac{2}{3})$ for $i\ge 2.$ We also measure an alternative quantity, the ratio of the BSM operators to the SM operator, which we define as \cite{Babich:2006bh},
\begin{equation}
R_i(\mu) =\bigg[ \frac{f_K^2}{m_K^2} \bigg]_{expt}\bigg[ \frac{m_K^2}{f_K^2}\frac{\braket{\bar{K}|O_i(\mu)|K}}{\braket{\bar{K}|O_1(\mu)|K}}\bigg]_{\mathrm{Lat.}}
\end{equation} At the continuum physical point, \hspace{0.01cm} $R_i(\mu)$ is a simple ratio of the BSM over the SM matrix element.

\section{Global Fit}

The physical value of $R_i(\mu)$ and $B_i(\mu)$ are recovered by performing a simultaneous chiral and continuum extrapolation. We first must renormalise the bare results. This was done non-perturbatively with non-exceptional kinematics (RI-SMOM) \cite{Sturm:2009kb} for $24^3$ and $32^3$ in ref. \cite{Garron:2016mva}. In these preliminary results, we have reused the renormalisation constants calculated for the $24(32)^3$ ensembles to renormalise $48(64)^3$. Next we adjust the valence strange mass to the physical point. Here there are multiple strange masses, we interpolate linearly. If there is only one strange mass for an ensemble, we extrapolate, taking the gradient from an ensemble set with the same lattice spacing and multiple strange masses.We ignore the size of the sea strange mass effects as they are expected to be negligible compared to the final error. These were measeured in \cite{1411.7017} Thirdly, the simultaneous fit is preformed, as in ref. \cite{Garron:2016mva}, according to one of two methods:
\begin{enumerate}
\item A fit to a linear ansatz:
\begin{equation}
Y\big(0,{m_\pi^2}/{(4\pi f_\pi)^2}\big) = Y(a^2,m^2_{ll}/(4\pi f_\pi))\bigg[1+c_a a^2 + c_m \frac{m_{ll}^2}{(4\pi f_{ll})^2}\bigg].
\end{equation}
\item  A fit according to NLO SU(2) chiral perturbation theory:
\begin{equation}
Y\big(0,{m_\pi^2}/{(4\pi f_\pi)^2}\big) = Y(a^2,m^2_{ll}/(4\pi f_{ll}))\bigg[1+c_a a^2 + \frac{m_{ll}^2}{(4\pi f_{ll})^2} ( c_m + C_i log(\frac{m_{ll}^2}{\Lambda^2})) \bigg].
\end{equation}
\end{enumerate} $a$ is the lattice spacing in $GeV^{-1}$, $m_{ll}$ is the pseudoscalar light-light meson mass. $c_a$ and $c_m$ are free parameters. The fixed coefficients $C_i$ are given in \cite{Becirevic:2004qd}. We then extrapolate to the physical point.

\section{Results}

\begin{figure}
	\centering
\begin{subfigure}{0.48\linewidth}
\includegraphics[width=\linewidth]{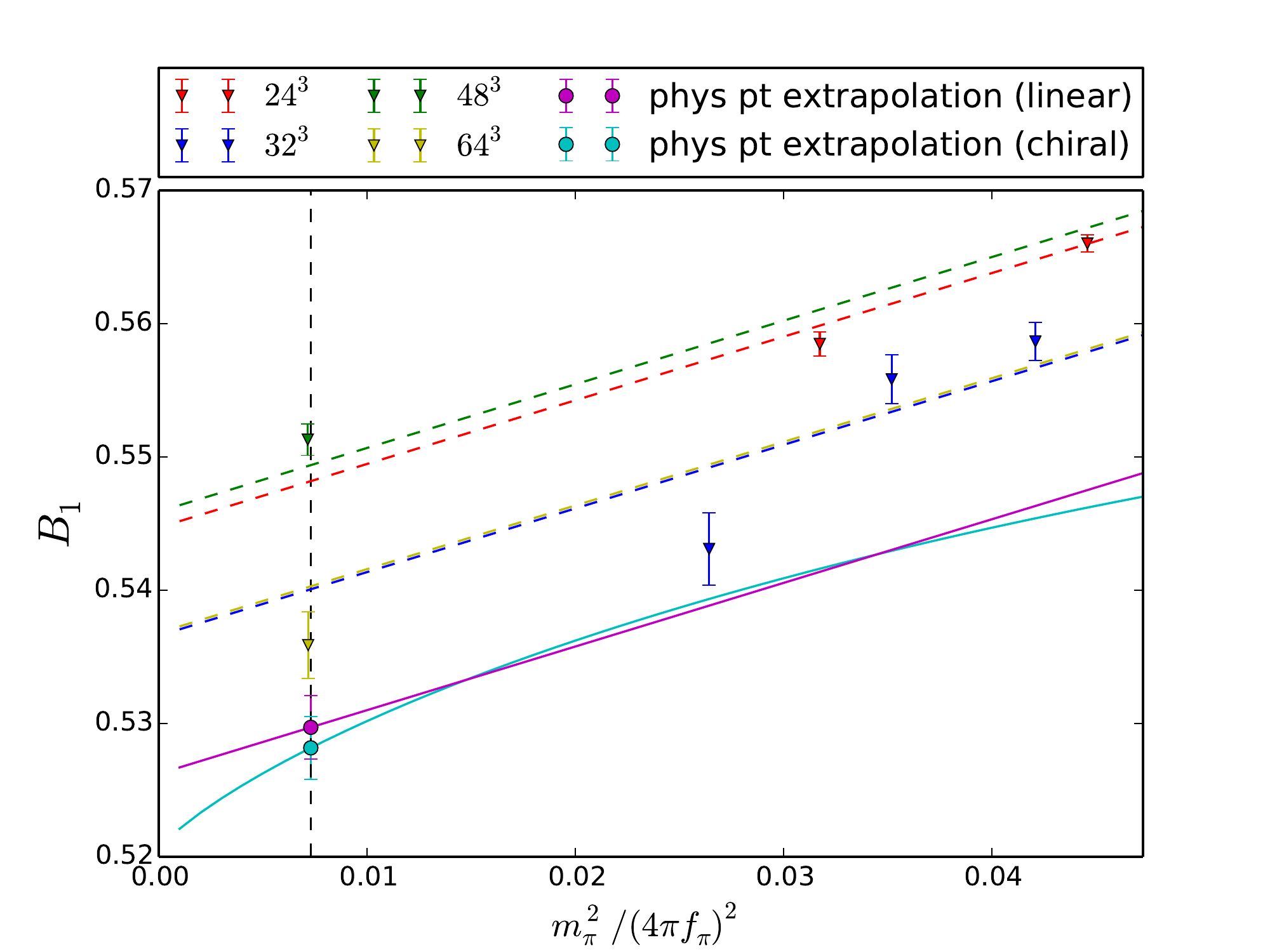}
\caption{$B_1^{\overline{ms}}(3GeV)$}
\end{subfigure}	
\begin{subfigure}{0.48\linewidth}
	\includegraphics[width=\linewidth]{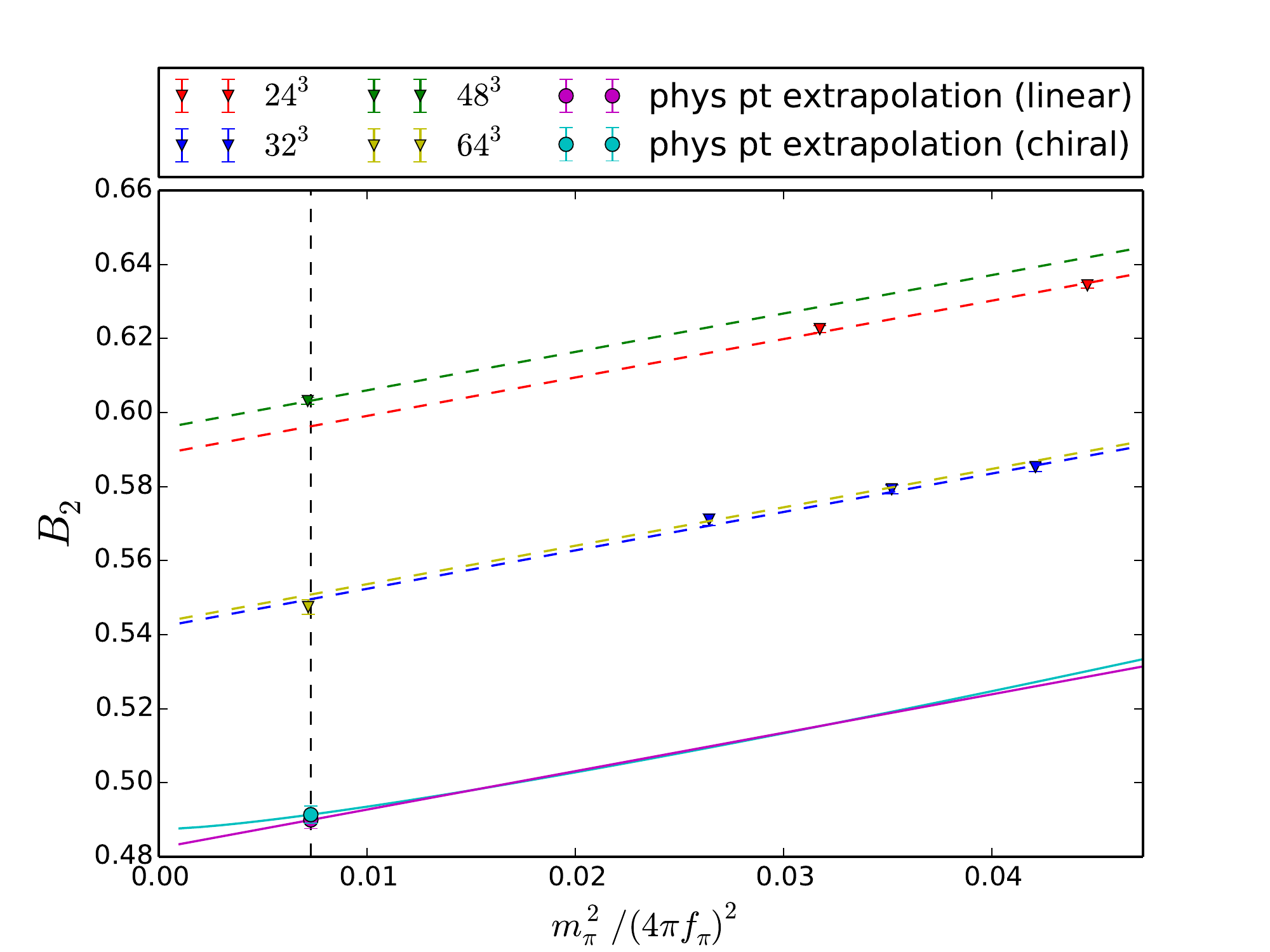}
\caption{$B_2^{\overline{ms}}(3GeV)$}
\end{subfigure}
\begin{subfigure}{0.48\linewidth}
	\includegraphics[width=\linewidth]{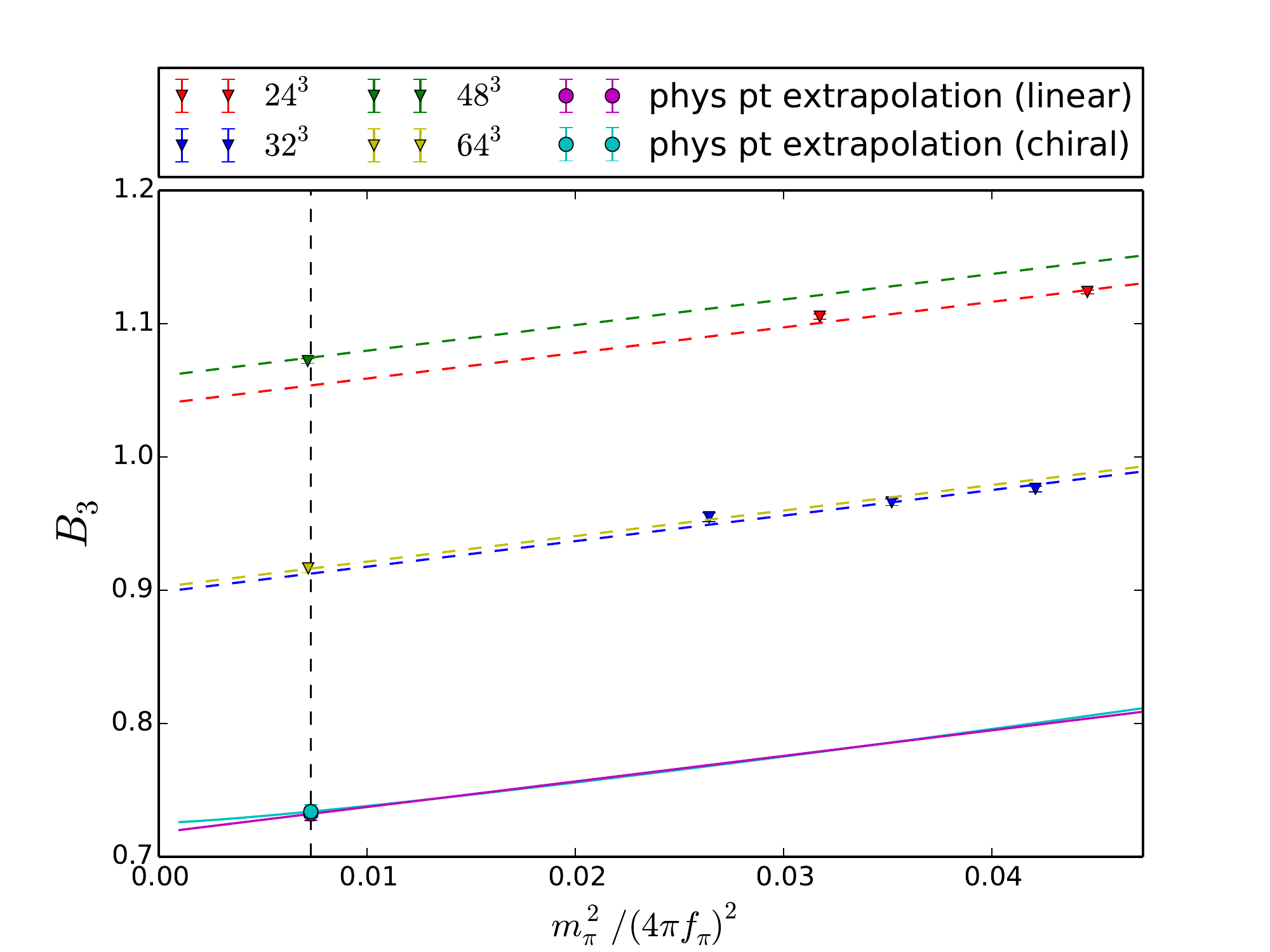}
\caption{$B_3^{\overline{ms}}(3GeV)$}
\end{subfigure}
\begin{subfigure}{0.48\linewidth}
	\includegraphics[width=\linewidth]{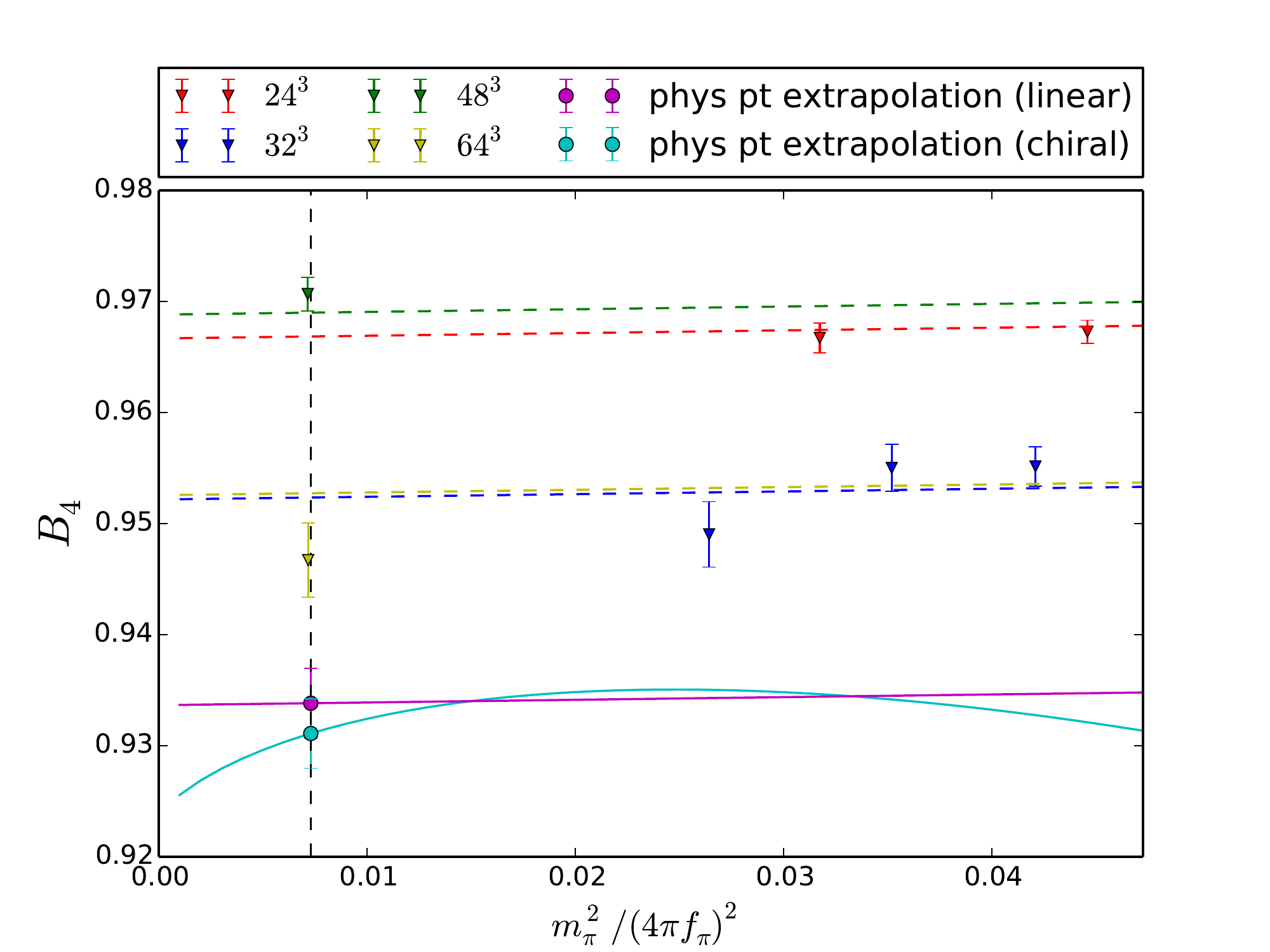}
\caption{$B_4^{\overline{ms}}(3GeV)$}
\end{subfigure}
\begin{subfigure}{0.48\linewidth}
	\includegraphics[width=\linewidth]{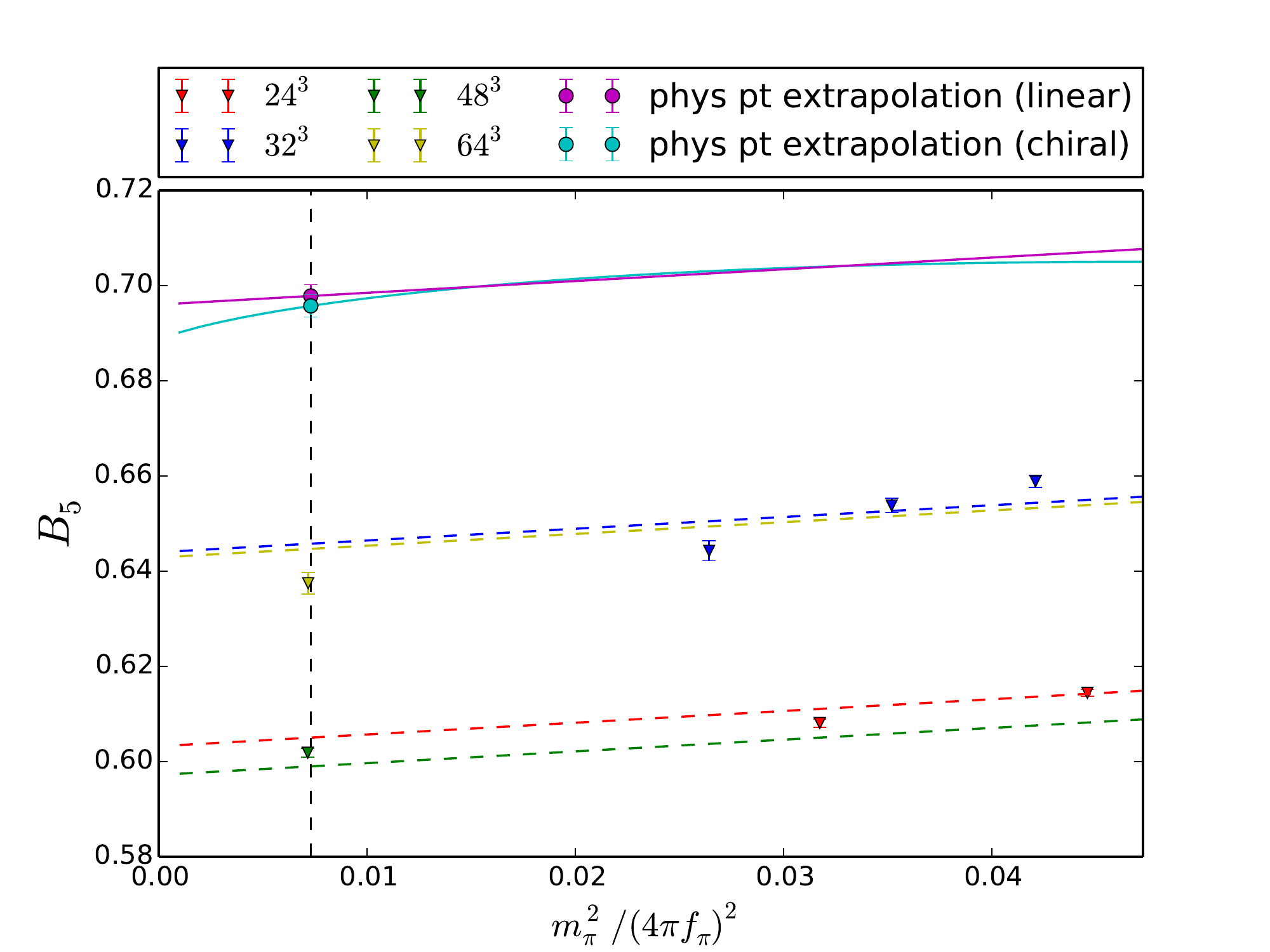}
\caption{$B_5^{\overline{ms}}(3GeV)$}
\end{subfigure}
\caption[Preliminary fits for the Bag parameters.]{
	Preliminary global fits for $B_i$. The linear fit lines for each lattice are shown. The physical point, continuum extrapolation result  and fit line at the continuum for each fit type are shown.}
\label{fig:Bs}

\end{figure}

\begin{figure}
	\centering
	\begin{subfigure}{0.48\linewidth}
		\includegraphics[width=\linewidth]{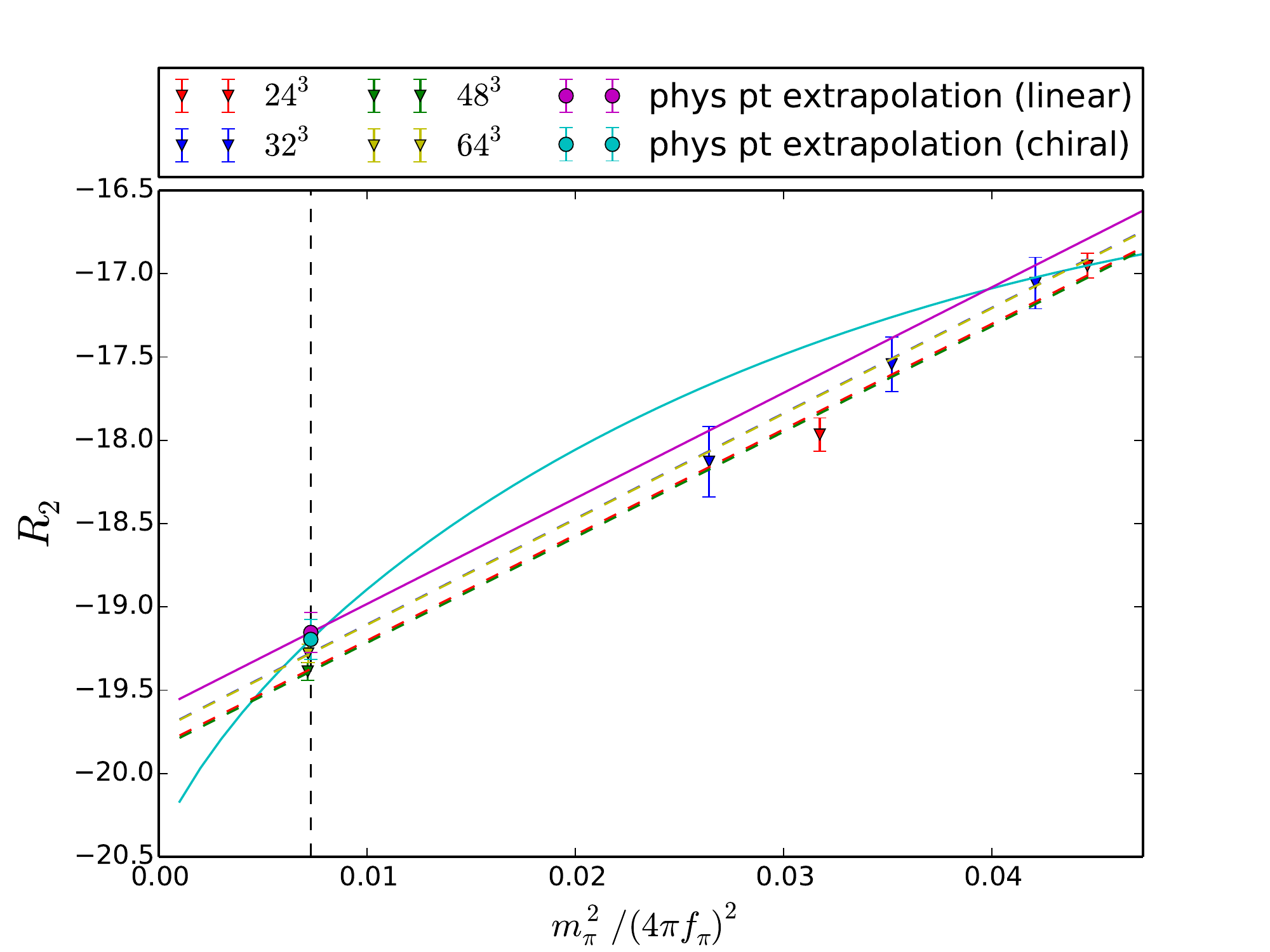}
		\caption{$R_2^{\overline{ms}}(3GeV)$}
	\end{subfigure}
	\begin{subfigure}{0.48\linewidth}
		\includegraphics[width=\linewidth]{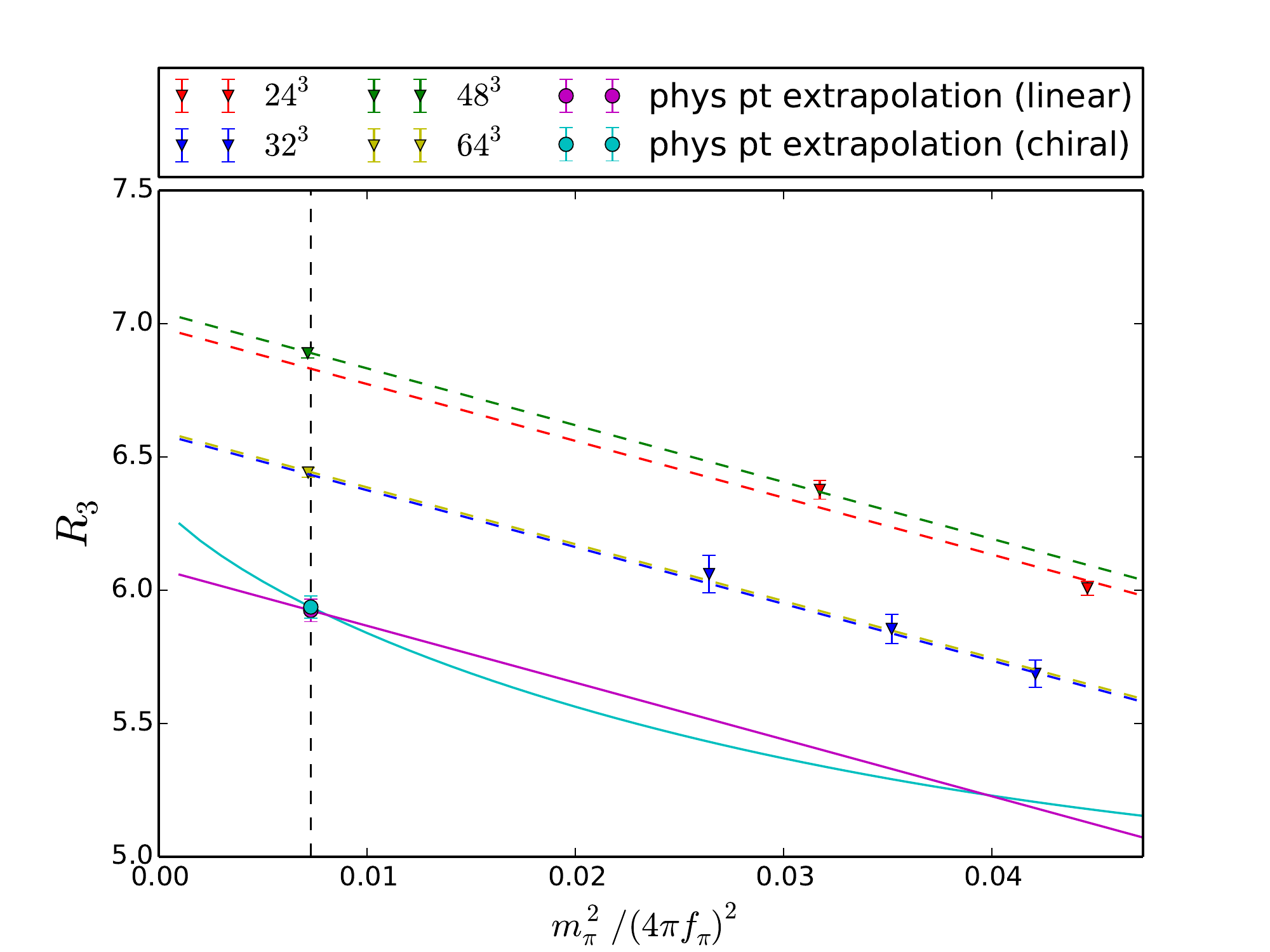}
		\caption{$R_3^{\overline{ms}}(3GeV)$}
	\end{subfigure}
	\begin{subfigure}{0.48\linewidth}
		\includegraphics[width=\linewidth]{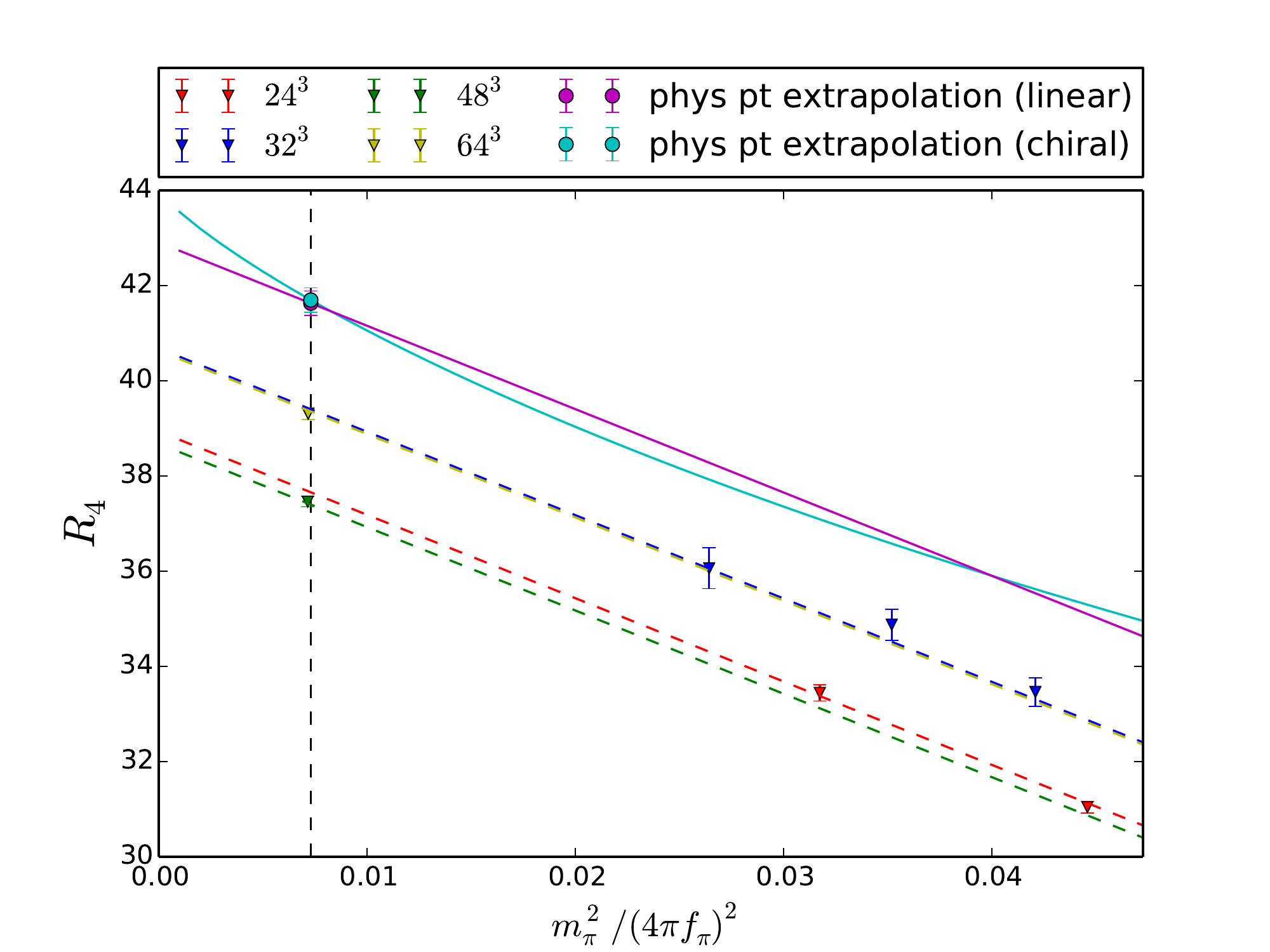}
		\caption{$R_4^{\overline{ms}}(3GeV)$}
	\end{subfigure}
	\begin{subfigure}{0.48\linewidth}
		\includegraphics[width=\linewidth]{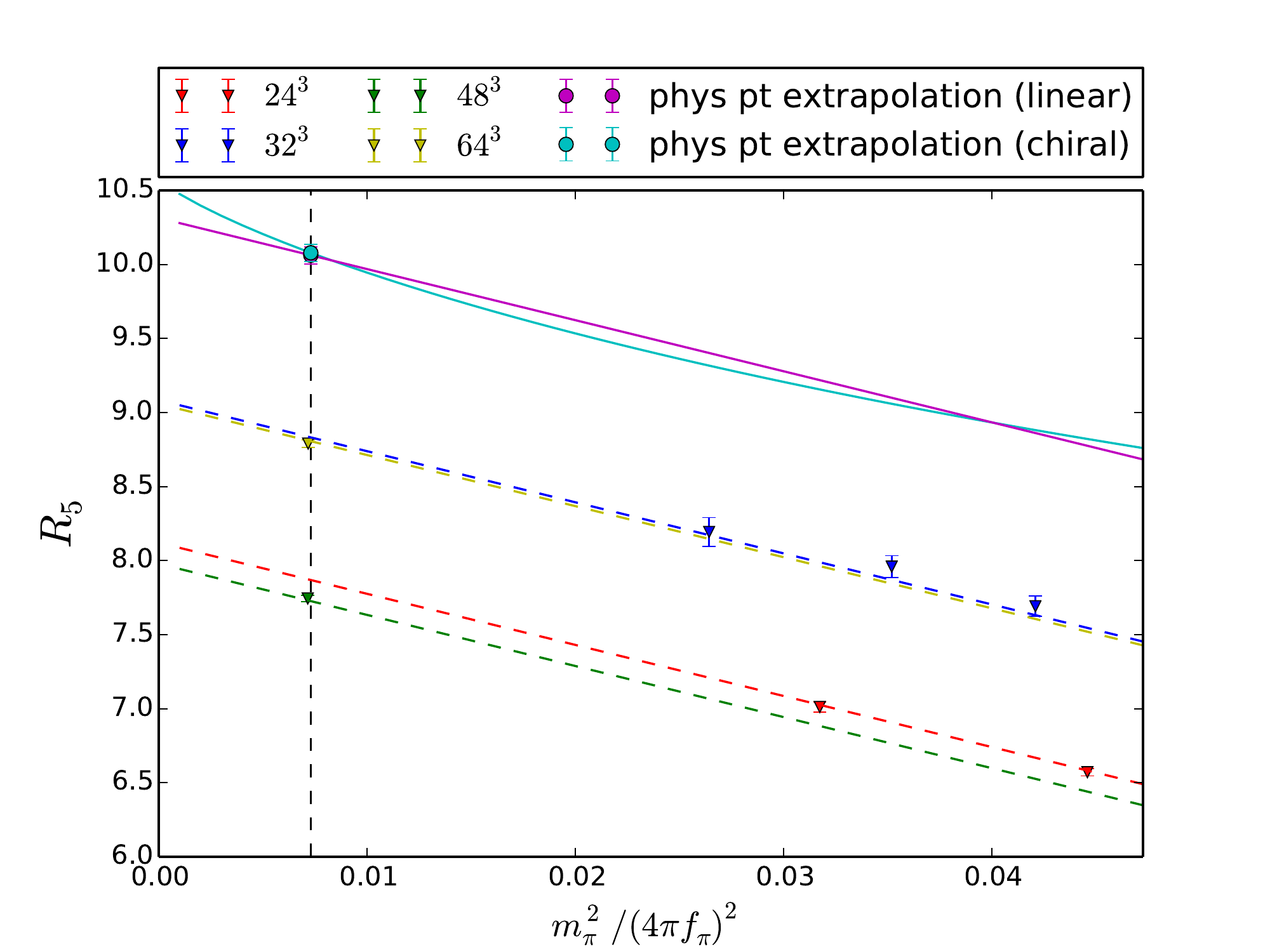}
		\caption{$R_5^{\overline{ms}}(3GeV)$}
	\end{subfigure}
\caption[Preliminary fits for the ratios.]{
	Preliminary global fits for $B_i$. The linear fit lines for each lattice are shown. The physical point, continuum extrapolation result  and fit line at the continuum for each fit type are shown.}
\label{fig:Rs}
\end{figure}

We present here results for the ratios and bag parameters. 
In table \ref{tab:raw_results} the bare unrenormalised values gained from the correlator fits are shown. We then present the preliminary continuum, physical point results gained from our fitting procedure in table \ref{tab:contres}.\footnote{It should be noted that all the errors presented on the new results here, are statistical errors only, and we are yet to calculate the systematic errors. We also have not shown here the errors arising from the matching to $\overline{ms}$ but these will be included in the future.} Plots of the fits are shown in figures \ref{fig:Bs} and \ref{fig:Rs}. We compare these new results to those previously gained by our collaboration. In \cite{1411.7017} a result for the SM bag parameter including the physical point data was found,  $B_k^{\overline{ms}}(3GeV) = 0.5293 \pm 0.0017 \pm 0.0106 \label{eq:old_phys}.$ This is consistent with the new result, even though we did not use overweighting of the physical point data and did not include reweighting for the sea strange mass. This agreement validates our statement that we can ignore these effects for the less well determined BSM matrix elements. These new results are (although  preliminary, and still to be improved upon) compatible with our collaborations previous BSM results \cite{Garron:2016mva} (shown in table \ref{tab:contres}) which were obtained with the same fitting approach.

\section{Conclusions}

We present preliminary results for the BSM bag parameters and ratios including physical pion mass data which appear to be consistent with the previous results in ref. \cite{Garron:2016mva}. We have dramatically reduced the chiral error on our results by including lattices with physical pions and kaons in our fits. We will include data from an ensemble with fine lattice spacing in our analysis for the final results.


\section{Acknowledgements} 
The ensemble results were computed on the STFC funded DiRAC BG/Q system in the Advanced Computing Facility at the University of Edinburgh. N.G. is supported by Leverhulme Trust, grant RPG-2014-118. The research leading to these results has received funding from the European Research Council under the European Union’s Seventh Framework Programme (FP7/2007-2013) / ERC Grant agreement 279757.
We are grateful to A. Khamseh for her contribution to calculating NPR for the fine ensemble which will be included in the final analysis.

\clearpage

\clearpage


	

\begin{table}[H]
	\centering
	\begin{tabular}{c | c c c c c }
		& $B_1$ & $B_2$ & $B_3$ & $B_4$ & $B_5$  \\
		\hline
		$48^3$ & 0.5840(9) & 0.9198(8) & 0.5767(5) & 0.9331(8) & 0.4951(4) \\
		$64^3$ & 0.5612(26)& 0.9082(30)  & 0.5282(19) & 0.9256(31)& 0.4488(16) \\		
		\hline\hline
		& - & $R_2$ & $R_3$ & $R_4$ & $R_5$ \\ 
		\hline
		$48^3$ & - & -23.596(67)& 35.903(103) & -18.489(51) &  -9.524(26) \\
		$64^3$ & -  & -27.054(69) &  41.426(107) & -20.046(47) & -10.149(24) \\
	\end{tabular}
	\caption{The  results of the bare bag parameters and ratios measured in this work, on the physical point ensembles, $m_\pi = 135.0 \: MeV$,  $m_K = 495.7 \:MeV.$}
	\label{tab:raw_results}
\end{table}
\begin{table}[H]
	\centering
	\begin{adjustbox}{width=\textwidth}  
		\begin{tabular}{ c c | c c c c c}
		& fit & $B_1$ & $B_2$ & $B_3$ & $B_4$ & $B_5$ \\
		\hline
		\multirow{2}{*}{new} & linear & 0.5320(4) & 0.4929(22) & 0.7388(49) & 0.9369(31) & 0.6999(24) \\
	 & chiral & 0.5305(23) & 0.4943(22) & 0.7405(49) & 0.9343(31) & 0.6978(24) \\
	 \hline
		prior work \cite{Garron:2016mva} & - & 0.536(9)(6)(11) & 0.492(7)(17)(5) & 0.751(14)(66)(8) & 0.932(12)(17)(13) & 0.680(8)(37)(26) \\
		\hline
		\hline
		 &  & - & $R_2$ & $R_3$ & $R_4$ & $R_5$ \\
		 \hline
		 \multirow{2}{*}{new} & linear &  -  & -19.10(11) & 5.928(37) & 41.28(23) & 9.989(5) \\
		 & chiral & - & -19.12(11) & 5.933(37) & 41.31(23) & 9.996(5) \\
		 \hline
		 prior work \cite{Garron:2016mva} & - & - & -19.91(45)(30)(43) & 6.22(16)(20)(14) & 44.25(91)(202)(115) & 10.68(20)(82)(31) \\ 
	\end{tabular}
	\end{adjustbox}
	\caption{Physical point continuum results obtained from a global fit for the $B_i$s and $R_i$s presented at $\overline{ms}(\mu=3\mathrm{GeV})$ with intermediate renormalisation  scheme RI-SMOM$^{\slashed{q}\slashed{q} }.$}
\label{tab:contres}
\end{table}

\end{document}